\documentstyle[preprint,aps]{revtex}
\begin{document}
%
%
\title{Indirect Constraints on the Triple Gauge Boson Couplings 
from $Z \to b \bar{b}$ Partial Width: An Update.}
\author{O.\ J.\ P.\ \'Eboli, M.\ C.\ Gonzalez--Garcia, and S.\ F.\ Novaes}
\address{Instituto de F\'{\i}sica Te\'orica,
Universidade Estadual Paulista, \\
Rua Pamplona 145, 01405--900 S\~ao Paulo, Brazil.}
\date{\today}
\maketitle

\begin{abstract}
We update the indirect bounds on anomalous triple gauge couplings
coming from the non--universal one--loop contributions to the $Z
\rightarrow b \bar{b}$ width. These bounds, which are independent
of the Higgs boson  mass, are in agreement with the standard
model predictions for the gauge boson self-couplings since the
present value of $R_b$ agrees fairly well with the theoretical
estimates. Moreover, these indirect constraints on $\Delta g^Z_1$
and $g^Z_5$ are more stringent than the present direct bounds on
these quantities, while the indirect limit on $\lambda_Z$ is
weaker than the available experimental data.
\end{abstract}

\newpage
%

\section{Introduction}

The predictions of the standard model of electroweak interactions (SM) agree
extremely well with the available experimental data \cite{ichep98}. The LEP I
and SLC collaborations studied in great detail the couplings of the $Z$ to
fermions, validating the $SU(2)_L\times U(1)_Y$ invariant interactions between
fermions and gauge bosons at the level of 1\%. At LEP II and at Tevatron we
are starting to probe the triple couplings among the weak gauge bosons with a
precision of the order of 10\% \cite{present}.  This is an important test of
the SM since these couplings are completely determined by the non--abelian
gauge symmetry of the model.

The $Z \to b \bar{b}$ width receives non--universal one--loop corrections due
to the presence of heavy particles running in the loop \cite{del:eps}, and it
is an important source of information on new physics beyond the standard
model. After the $R_b$ crisis has been solved, it is important to revisit all
the phenomenological analyses that are strongly based on this quantity. In
particular, the precise measurement of the $Z \to b \bar{b}$ width is able to
constrain possible deviations of the triple gauge--boson couplings with an
accuracy that, in some cases, are even better than the direct measurements of
these interactions. In this letter, we make an update on our previous analysis
taking into account recent experimental data on the electroweak parameters. We
present our results in terms of effective Lagrangians for the anomalous gauge
interactions both for linear as well as non-linear realization of the $SU(2)_L
\times U(1)_Y$ symmetry \cite{deruj,Appelquist}.

The non--universal contributions to the $Z b \bar{b}$ couplings have been
parametrized, in a model independent way, in terms of the parameter
$\epsilon_b$ \cite{def:eb}, defined as
\begin{equation}
        \epsilon_b \equiv \frac{g^b_A}{g^\ell_A} - 1 \; ,
\end{equation}
where $g^b_A$ ($g^\ell_A$) is the axial coupling of the $Z$ to $b \bar{b}$
($\ell \bar{\ell}$) pairs. An important feature of this parameter is that its
SM value is basically independent of the Higgs boson mass. Therefore, the
bounds withdrew from it have less uncertainties. The anomalous contribution to
the $Z f \bar{f}$ vertex can be written in terms of the form factor $F(m_j)$
as,
\begin{equation}
\Gamma_{\text{ano}}^\mu (Zf\bar{f}) =  i\frac{e}{4 s_W c_W}
\sum_i V_{if} V^\dagger_{if} F(m_j) \gamma^\mu (1-\gamma^5) \; ,
\end{equation}
where $V_{if}$ is the Cabibbo--Kobayashi--Maskawa mixing matrix in the case of
quarks and $V_{if} = \delta_{if}$ for leptons.  This amplitude is the same for
all external fermions but the $b$ quark when we neglect the mixings
$V_{td(s)}$ and all the internal fermions masses but $m_{\text{top}}$.
Therefore, $\epsilon_b$ takes the form
\[
\epsilon_b = \Delta F \equiv   F(m_{\text{top}}) - F(0)   \; .
\]
The contribution of anomalous $W^+W^-Z$ couplings to the $Z\rightarrow b \bar
b$ partial width was evaluated previously in Refs.\ 
\cite{our:epsb,other:epsb}. In this process, the non--universal effect are
enhanced, as expected, by powers of the top quark mass due to the virtual top
quark running in the loop vertices corrections \cite{del:eps}.

\section{Effective Lagrangians}

The usual Lorentz invariant and CP conserving pa\-ra\-metrization of the
$W^+W^-V$ vertex, with $V = \gamma$ or $Z$, is given by the effective
Lagrangian \cite{old:wwv},
\begin{eqnarray}
{\cal L}^{\text{WWV}}_{\text{eff}} =&&  - i g_{\text{WWV}}
\left [
g^V_1 ( W^+_{\mu\nu} W^{-\mu} - W^-_{\mu\nu} W^{+\mu} ) V^\nu \right.
\nonumber \\
&& + \kappa_V W^+_\mu W^-_\nu V^{\mu\nu} +
\frac{\lambda_V}{M_W^2} W_\mu^{+\nu} W_\nu^{-\rho} V_\rho^\mu   
\nonumber \\
&& \left. - i g^V_5 \epsilon^{\mu\nu\rho\sigma}
(W^+_\mu \partial_\rho W^-_\nu
- W^-_\nu \partial_\rho W^+_\mu ) V_\sigma  \right ]
\label{WWV}
\end{eqnarray}
where $V_{\mu\nu} = \partial_\mu V_\nu - \partial_\nu V_\mu$, $g_{WW\gamma} =
e$, and $g_{\text{WWZ}} = e c_W/s_W$, with $s_W (c_W) =\sin (\cos) \theta_W$.
The first three terms in Eq.\ (\ref{WWV}) are C and P invariant while the last
one violates both C and P.

Since the standard model is consistent with the available experimental data,
it is natural to parametrize the anomalous triple gauge boson couplings in
terms of an effective Lagrangian which exhibits the $SU(2)_L\times U(1)_Y$
gauge invariance. The particular way this symmetry is realized depends on the
particle content at low energies. If a light Higgs boson is present, the
symmetry can be realized linearly \cite{deruj,hisz}, and the leading effects
of new interactions are described by eleven dimension--6 operators ${\cal
  O}_i$
\begin{equation}
{\cal L}^{\text{linear}}_{\text{eff}}  = 
\sum_i \frac{f_i}{\Lambda^{2}} {\cal O}_i  \; ,
\end{equation}
at energies below the new physics scale $\Lambda$. Three of these
operators \cite{deruj}, namely,
\begin{eqnarray}
{\cal O}_{B}   &=& (D_\mu \Phi)^\dagger
\hat B^{\mu\nu} (D_\nu \Phi) \; , 
\nonumber \\
{\cal O}_{W} &=& (D_\mu \Phi)^\dagger
\hat W^{\mu\nu} (D_\nu \Phi)  \; ,
\label{lin:op}
\\ 
{\cal O}_{WWW} &=& {\text Tr} \left [\hat W_{\mu\nu}
\hat W^{\nu\rho}\hat W_\rho^{\mu} \right ] \; , 
\nonumber
\end{eqnarray}
modify the triple gauge boson couplings without affecting the gauge boson
two--point functions at tree level (``blind'' operators). In our notation,
$\hat{B}_{\mu\nu} = i (g^\prime/2) B_{\mu\nu}$ and $\hat{W}_{\mu\nu} = i (g/2)
\sigma^a W^a_{\mu\nu}$ with $B_{\mu\nu}$ and $W^a_{\mu\nu}$ being the $U(1)_Y$
and $SU(2)_L$ full field strengths and $\sigma^a$ representing the Pauli
matrices. In this framework, it is expected that $g^Z_5$ should be suppressed
since it is related to a dimension 8 operator \cite{our:epsb}.

The anomalous couplings of the parametrization (\ref{WWV}) are related to the
coefficients of the linearly realized effective Lagrangian by
\begin{eqnarray}
\Delta g^Z_1 &=& f_W~\frac{m_Z^2}{2\Lambda^2} \; ,
\label{lin:1} \\
\Delta \kappa_Z &=& [f_W-s_W^2 (f_B+f_W)]~\frac{m_Z^2}{2\Lambda^2} \; ,
\label{lin:2} \\
\lambda_Z &=& f_{WWW}~ \frac{3m_W^2 g^2}{2\Lambda^2} \; , 
\label{lin:3}
\end{eqnarray}
where $\Delta \kappa_V \equiv \kappa_V - 1$, $\Delta g_1^Z \equiv g_1^Z - 1$,
and $\lambda_V$ are all zero in the SM at tree level.  It is interesting to
notice that these effective operators lead to the following relation between
the coefficients of Lagrangian (\ref{WWV}):
\begin{eqnarray}
        \Delta \kappa_\gamma &=& \frac{c_W^2}{s_W^2} 
        \left( \Delta g_1^Z - \Delta \kappa_Z  \right)
        \; , \label{rel1}
\\
        \lambda_\gamma &=& \lambda_Z \; .
\label{rel}
\end{eqnarray}

In the scenario where the $SU(2)_L \times U(1)_Y$ gauge symmetry is
non--linearly realized \cite{Appelquist}, a chiral Lagrangian can be
constructed from the dimensionless unitary matrix $U$ that belongs to the
$(2,2)$ representation of the group $SU(2)_L \times SU(2)_C$,
\begin{equation}
{\cal L}^{\text{non--lin.}}_{\text{eff}}  = 
\sum_i  \alpha_i {\cal O}_i  \; .
\end{equation} 
The ``blind'' directions that appear in the lowest order of the chiral
expansion are described by the Lagrangians \cite{Appelquist},
\begin{eqnarray}
{\cal O}_{2} &=& \frac{i g^\prime}{2}  B^{\mu\nu} \text{Tr} 
\left ( T \, [(D_\mu U) U^\dagger, (D_\nu U) U^\dagger] \right )
\; , \nonumber \\
{\cal O}_{3} &= &\frac{i g}{2} \text{Tr}
\left ( W^{a\mu\nu}\sigma^a 
T \, [(D_\mu U) U^\dagger, (D_\nu U) U^\dagger] \right )
\; , \label{nonlin:op} \\ 
{\cal O}_{9} &=& \frac{i g}{4}
\text{Tr}\left ( T W^{a\mu\nu}\sigma^a \right )
\text{Tr}\left ( T [ (D_\mu U) U^\dagger, (D_\nu U) U^\dagger] \right )
\; , \nonumber \\
{\cal O}_{11} &=& \frac{g}{2} \epsilon^{\mu\nu\lambda\rho}
\text{Tr}\left [T (D_\mu U) U^\dagger \right ]
\text{Tr}\left [ (D_\nu U) U^\dagger W^a_{\lambda\rho} \sigma^a
\right] \nonumber \; .
\end{eqnarray}
where the custodial symmetry breaking operator $T \equiv U \sigma_3 U^\dagger$
and the covariant derivative of $U$ is defined as $D_\mu U \equiv \partial_\mu
U + i (g/2) \sigma^a W^a_\mu U - i (g^\prime/2) U \sigma_3 B_\mu$.

The contribution of the above chiral operators can be expressed in terms of
the standard parametrization as \cite{Appelquist}
\begin{eqnarray}
\label{rel:chi}
\Delta g_1^Z &=& \alpha_3 \frac{g^2}{c_W^2} \; ,
\label{nlin:1} \\
\Delta \kappa_Z &=& \left [ c_W^2(\alpha_3 + \alpha_9 )-  s_W^2 \alpha_2
\right ] \frac{g^2}{c^2_W} \; , 
\label{nlin:2} \\
g_5^Z &=& \alpha_{11} \frac{g^2}{c_W^2} 
\; . \label{nlin:3} 
\end{eqnarray}


\section{Results and Conclusions}

The contribution of the anomalous couplings (\ref{WWV}) to $\epsilon_b$ can be
written as
\begin{eqnarray}
\epsilon_{b}^{\text{ano}} \equiv \epsilon_{b} - \epsilon_{b}^{\text{SM}} 
&=& 
\Delta\kappa_Z \Delta F_{\kappa_Z}  +
\Delta g_1^Z \Delta F_{g_1^Z}  
\nonumber \\ 
&+&  \lambda_Z \Delta F_{\lambda_Z}  + g_5^Z \Delta
F_{g_5^Z}
\label{epsiano}
\end{eqnarray}
where the form factors $\Delta F_{\kappa_Z}$, $\Delta F_{g_1^Z}$, $\Delta
F_{\lambda_Z}$, and $\Delta F_{\lambda_Z}$ were presented elsewhere
\cite{our:epsb}.

In order to obtain our numerical results we used the most recent data for the
electroweak parameters and masses \cite{ichep98}: $\alpha (M_Z) = 1/128.896$,
$s_W^2 = 0.2321$, $M_Z = 91.1867$ GeV, $M_W = 80.37$ GeV, and $m_{\text{top}}
= 173.8$ GeV.  Substituting these parameters into the expressions for the form
factor \cite{our:epsb}, we obtain
\begin{eqnarray}
\epsilon_{b}^{\text{ano}} &=& 
\Delta\kappa_Z \left[- 3.1 \times 10^{-3}
\log\left(\frac{\Lambda^2}{M_W^2} \right) \right]  
\nonumber \\
&+& \Delta g_1^Z  \left[- 1.4 \times 10^{-2}
\log\left(\frac{\Lambda^2}{M_W^2} \right) \right]   
\nonumber \\
&+& \lambda_Z \left( - 2.6 \times 10^{-3} \right)  
+ g_5^Z \left( - 8.3 \times 10^{-3} \right) \; .
\label{eb:ano}
\end{eqnarray}
The form factors $\Delta F_{\lambda_Z}$ and $\Delta F_{g^Z_5}$ are independent
of the cutoff $\Lambda$, while the form factors $\Delta F_{\kappa_Z}$ and
$\Delta F_{g_1^Z}$ are ultra-violet divergent which indicate a logarithmic
dependence in $\Lambda$.  Since the $\log \Lambda$ terms are dominant in these
form factors, we dropped the constant term from their expressions.

The SM prediction for $\epsilon_b$ is practically independent of the Higgs
boson mass, and for $m_{\text{top}} = 173.8$ GeV its value is
$\epsilon_b^{\text{SM}} = -6.51 \times 10^{-3}$ \cite{epsb:new}. On the other
hand, a global fit to the available data leads to $\epsilon_b^{\text{exp}} =
(-3.9 \pm 2.1)\times 10^{-3}$ \cite{epsb:new}.  The constraints on the
couplings of the effective Lagrangian (\ref{WWV}) can be easily obtained using
the SM and experimental values of $\epsilon_b$ and expression (\ref{eb:ano}).
We present in Table~\ref{tab} our 1-$\sigma$ limits on $\Delta g^Z_1$, $\Delta
\kappa_Z$, $\lambda_Z$, and $g_5^Z$, assuming that only one coupling at a time
is allowed to deviate from zero and taking $\Lambda = 1$ TeV.

At this point it is interesting to compare our indirect bounds with the
present direct limits on the anomalous triple gauge boson couplings.  Taking
into account both LEP and D\O ~data, the allowed range of the parameters
$\Delta \kappa_\gamma $, $\Delta g_1^Z$, and $\lambda_\gamma $ are
\cite{ichep98}
\begin{eqnarray}
\Delta \kappa_\gamma &=& 0.13 \pm 0.14 
\; , \nonumber \\
\Delta g_1^Z &=& 0.00 \pm 0.08
\; , \nonumber \\
\lambda_\gamma &=& -0.03 \pm 0.07 \;\; .
\nonumber
\end{eqnarray}
Therefore, our indirect bounds on $\Delta g^Z_1$ is a factor of 4 more
stringent than the present direct limit. Moreover, the above experimental
results assumed the $SU(2)$ invariant relations (\ref{rel1}) and (\ref{rel}).
Using this hypothesis, our indirect constraint on $\lambda_\gamma$
($=\lambda_Z$) turns out to be a factor of 15 looser than the available direct
bound.

The relations (\ref{lin:1})--(\ref{lin:3}) and (\ref{nlin:1})--(\ref{nlin:3})
allow us to derive bounds on the ``blind'' operators (\ref{lin:op}) and
(\ref{nonlin:op}). We also present our constraints on these couplings in
Table~\ref{tab}, where we assumed $\Lambda=1$ TeV and that only one coupling
is non--vanishing at a time.  For the sake of comparison we show here the
combined LEP limits on some of these operators
\begin{eqnarray*}
\frac{M_W^2}{2 \Lambda^2} f_W &=& -0.05 \pm 0.06 
\;\; , \\
\frac{M_W^2}{2 \Lambda^2} f_B &=& -0.04^{+0.33}_{-0.24} 
\;\; , \\
\frac{3 M_W^2 g^2}{2\Lambda^2} f_{WWW} &=& -0.09^{+0.13}_{-0.12} \;\; .
\end{eqnarray*}
As we can see, the indirect limits on $f_W$ and $f_B$ are of the same order of
the experimental ones while the direct bound on $f_{WWW}$ is much better. It
is interesting to notice that the indirect limits of operators, which lead to
divergent one--loop contributions to the vertex and consequently are enhanced
by factors $\log (\Lambda/M_W)$, are the only ones competitive with the
present experimental results.

It is also possible to constrain the triple gauge boson couplings via the
analysis of rare $B$ and $K$ decays \cite{low,gus}.  Recently, Burdman has
obtained the 1-$\sigma$ limits $|\Delta g_1^Z| < 0.10$ and $|\Delta
\kappa_\gamma| < 0.20$, for a new physics scale $\Lambda = 2$ TeV.  In order
to compare our results with this work we derive the 1-$\sigma$ bounds for
$\Lambda = 2$ TeV:
\begin{eqnarray*}
-0.051 < &&\Delta g_1^Z  < -0.0055 \;\; , 
\\
-0.24  < &&\Delta \kappa_Z < -0.026 \;\; ,
\end{eqnarray*}
when just one anomalous coupling is non-vanishing in each analyses.  Taking
into account the relation given in Eq.\ (\ref{rel1}), we can also derive an
indirect constraint on $\Delta \kappa_\gamma$ when just $\Delta g_1^Z$ is
different from zero, {\it i.e.\/} $-0.17 < \Delta \kappa_\gamma < 0.018$. This
shows that the limits obtained from the analysis of the data on $Z\rightarrow
b\bar b$ is in some cases, more than one order of magnitude better than the
one coming from the $B$ and $K$ decays.


\acknowledgments 
M.C. G-G is very grateful to the Instituto de F\'{\i}sica
Te\'orica da Universidade Estadual Paulista for their kind hospitality.  This
work was supported by Conselho Nacional de Desenvolvimento Cient\'{\i}fico e
Tecnol\'ogico (CNPq), by Funda\c{c}\~ao de Amparo \`a Pesquisa do Estado de
S\~ao Paulo (FAPESP), and by Programa de Apoio a N\'ucleos de Excel\^encia
(PRONEX).



\widetext

\begin{table}
\begin{tabular}{||c|c||c|c||c|c||}
\multicolumn{2}{||c||}{Eq.\ (\ref{WWV})}   & 
\multicolumn{2}{c||}{Eq.\ (\ref{lin:op})} & 
\multicolumn{2}{c||}{Eq.\ (\ref{nonlin:op})} \\
\hline\hline
$\Delta g^Z_1$        & $-0.036\pm 0.029$       
& $M_W^2/(2\Lambda^2) \; f_B$               & $0.56 \pm 0.45$  
& $\alpha_2$          & $1.3\pm 1.1$
\\ \hline
$\Delta\kappa_Z$       & $-0.17\pm 0.13$ 
& $M_W^2/(2\Lambda^2) \; f_W$               & $-0.024 \pm 0.019$  
& $\alpha_3$          & $-0.057\pm 0.046$       
\\ \hline
$\lambda_{\gamma,Z}$  & $-1.0 \pm 0.81$
& $(3 M_W^2 g^2)/(2 \Lambda^2) \; f_{WWW}$ & $ -1.0 \pm 0.81$ 
& $\alpha_9$          & $-0.40\pm 0.32$
\\ \hline
$g_5^Z$               & $-0.31\pm 0.23$ 
& &  
 & $\alpha_{11}$       & $-0.57\pm 0.46$ 
\end{tabular}
\caption{One-$\sigma$ allowed regions of the anomalous triple
gauge--boson  couplings in different parametrizations,  assuming
$\Lambda = 1$ TeV.}
\label{tab}
\end{table}

\end{document}